\documentclass[preprint,aps,floatfix,superscriptaddress]{revtex4}

\usepackage{dcolumn}
 \usepackage{ulem}

\usepackage{graphicx}

\usepackage{amsfonts}
\usepackage{amsmath,bm,amssymb}
\usepackage[usenames]{color}
\usepackage{comment}

\def\bfk{{\bf k}}

\begin{document}

\title{ Direct theoretical evidence for weaker correlations in
  electron-doped and Hg-based hole-doped cuprates}

\author{Seung Woo Jang}
\affiliation{Department of Physics, Korea Advanced Institute of
  Science and Technology (KAIST), Daejeon 305-701, Korea }

\author{Hirofumi Sakakibara} \affiliation{Computational Condensed
  Matter Physics Laboratory, RIKEN, Wako, Saitama 351-0198, Japan
}
  
\author{Hiori Kino} \affiliation{National Institute for Materials
  Science, Sengen 1-2-1, Tsukuba, Ibaraki 305-0047, Japan.}

\author{Takao Kotani} \affiliation{Department of Applied Mathematics
  and Physics, Tottori University, Tottori 680-8552, Japan}

\author{Kazuhiko Kuroki} 
\affiliation{Department of Physics, Osaka University,
  Machikaneyama-Cho, Toyonaka, Osaka 560-0043, Japan}

\author{Myung Joon Han} \email{mj.han@kaist.ac.kr}
\affiliation{Department of Physics, Korea Advanced Institute of
  Science and Technology (KAIST), Daejeon 305-701, Korea }
\affiliation{ KAIST Institute for the NanoCentury, Korea Advanced
  Institute of Science and Technology, Daejeon 305-701, Korea }

\begin{abstract}
Many important questions for high-$T_c$ cuprates are closely related to 
the insulating nature of parent compounds.
  While
there has been intensive discussion on this issue, all
arguments rely strongly on, or are closely related to, the correlation
strength of the materials. Clear understanding has been seriously
hampered by the absence of a direct measure of this interaction,
traditionally denoted by $U$. Here, we report a first-principles
estimation of $U$ for several different types of cuprates.  The $U$
values clearly increase as a function of the inverse bond distance
between apical oxygen and copper. Our results show that the
electron-doped cuprates are less correlated than their hole-doped
counterparts, which supports the Slater picture rather than
the Mott picture.  Further, the $U$ values significantly vary even
among the hole-doped families.  The correlation strengths of the
Hg-cuprates are noticeably weaker than that of La$_2$CuO$_4$. 
Our results suggest that
the strong correlation enough to induce Mott gap may not be a prerequisite
for the high-$T_c$ superconductivity.
\end{abstract}

\pacs{74.20.Pq, 74.72.Ek, 74.25.Dw, 71.15.-m}

\maketitle

 Due to extensive efforts over the last 30 years
\cite{first1986}, significant progress has been made in the
understanding of high-temperature superconducting materials. Although
the pairing mechanism and the intriguing interplay between competing
orders still remain elusive, many aspects of this series of
copper-oxides have now been well established.  Basically, all cuprates
share common phase diagram features, and each phase has been a subject
of intensive study. The `dome'-shaped region of superconductivity,
which only appears after the long-range magnetic order is suppressed
(see Figure 1), is possibly the key to understanding the pairing
principle of cuprates. These features are also found in other families
of superconducting materials, such as Fe-based and heavy Fermion
compounds, and have been well recognized, likely suggesting that the
same superconducting mechanism exists in the different families
\cite{review18}.

The superconducting dome has been considered to be particularly
important in the framework of some outstanding theoretical models or
`pictures' that assume or predict its existence
\cite{review10,review12}. Therefore, it is striking that a series of
recent experiments for electron-doped cuprates have reported data that
contradicts this feature. According to a systematic re-investigation
of electron-doped samples, RE$_2$CuO$_4$ (RE=rare-earth: Nd, Pr, Sm,
etc.), the superconducting region does not cease to exist as the
carrier concentration decreases, but this region extends to very low
doping, quite close to zero
\cite{Brinkmann-PRL1995,Matsumoto-PC2008,Matsumoto-PRB2009,Matsumoto-PC2009,
  Matsumoto-PC2009-2,Matsumoto-PC2010,Yamamoto-SSP2011,Krockenberger-PRB2012,
  Krockenberger-SR2013,Chanda-PRB2014,Tsukada-SSC2005,Adachi-JPSJ2013,
  review17,review-PhysicaC}. Further, as the doping approaches zero,
the superconducting transition temperature ($T_c$) seems to keep
increasing with no indication of the dome (see Figure 1(b)).
While further study needs to be performed to clarify this issue,
it seems indicative that the undoped parent compounds of
RE$_2$CuO$_4$ are a Slater-type insulator rather than a Mott-type
insulator. Therefore, the `doped Mott insulator' picture may not be
appropriate, at least for the electron-doped family.

Some theoretical suggestions are supportive of this conclusion.
According to Weber {\it et al.}  \cite{Weber-NP2010,Weber-PRB2010},
for example, an electron-doped material, Nd$_2$CuO$_4$, is less
correlated and should be identified as a Slater insulator, while the
hole-doped La$_2$CuO$_4$ should be considered as a Mott insulator. The
LDA+DMFT (local density approximation plus the dynamical mean field
theory) calculation by Das and Saha-Dasgupta \cite{Das-PRB2009} showed
that the $T$-structured La$_2$CuO$_4$ is insulating while the
$T'$-structured La$_2$CuO$_4$ is metallic at $U$ = 4.5 eV.  Comanac
{\it et al.} \cite{Comanac-NP2008} also concluded that the correlation
strengths in cuprates are not strong enough to be identified as Mott
insulators.

In spite of its crucial importance, however, this issue is quite
challenging because of the difficulty in quantifying the `Mott-ness'
or in estimating the correlation strengths. Here, we also note that
while Comanac {\it et al.}  concluded that all the cuprates are Slater
insulators \cite{Comanac-NP2008}, Weber {\it et al.}, as well as Das
and Saha-Dasgupta, made a sharp distinction between the electron-doped
and the hole-doped families
\cite{Weber-NP2010,Weber-PRB2010,Das-PRB2009}.  One clear and
well-defined way for resolving this issue is to calculate or `measure'
the material dependence of the correlation strength, which is
traditionally denoted by the parameter $U$ (on-site Coulomb repulsion
within the single-band Hubbard model).  Further, calculating the
material dependent $U$ values can illuminate other important issues
such as pairing principle.  Because electron-doped cuprates generally
have lower $T_c$ ($\leq$ 30 K) than hole-doped materials, whose $T_c$
sometimes exceeds 100 K ({\it e.g.}, the triple-layered Hg-cuprates),
it is important to determine if there is a notable difference in the
correlation strengths of these two different families.

Here, we try to provide a clear answer to this long standing
question by performing the direct estimation of $U$ for several
different types of cuprates.  Our first-principles calculations show
that both of the previous conclusions are not quite correct. 
On one
hand, our result provides the first direct confirmation that the
correlation strength of electron-doped materials is weaker than that
of hole-doped counterparts. On the other, we significantly revise the
previous conclusion: Not all of the hole-doped cuprates 
have stronger correlation compared to the electron doped ones.
In fact, one representative hole-doped family, namely
Hg-cuprates (and presumably many other multi-layered cuprates), has
weaker electron correlation strength comparable to the electron-doped
materials.   
Our result has a profound implication for the pairing principle: The
correlation effects, strong enough to produce the Mott insulating
state, may not be a prerequisite for high $T_c$ superconductivity.

\section*{Results}

 The results are summarized in Figure
2. We clearly see that T'-structures (or, the parent compounds of electron-doped materials)
 have significantly
smaller $U$ values than the hole-doped materials (parent phases), especially
La$_2$CuO$_4$. The calculated $U$ for RE$_2$CuO$_4$ (RE: Nd, Pr, Sm)
is 1.24--1.34 eV, which is considerably smaller than the La$_2$CuO$_4$
value of 3.15 eV. The material dependent $U/t$ was also estimated (see
Figure 2; the data in green color and the right vertical axis), where
the nearest-neighbor hopping parameter, $t$, was calculated with the
standard Wannier-function technique
\cite{Marzari-PRB1997,Souza-PRB2001} (see Supplementary Information). The calculated $U/t$ for
La$_2$CuO$_4$ is $\sim$7 which compares reasonably well with the
widely used values for the model Hamiltonian studies
\cite{Araujo-EPL2012}.  The $U/t$ value for the RE$_2$CuO$_4$ series
is $\sim$3, which is significantly smaller 
($\sim$43\% of the La$_2$CuO$_4$ value).

The $4f$ electrons in RE$_2$CuO$_4$ located around the Fermi level
must be considered carefully. Because there is no well-established
method to treat these states, first-principles calculations of
rare-earth compounds has been challenging. One widely-used method is
to treat the $4f$ electrons as part of the core electrons, as was done
in Ref.~\onlinecite{Weber-NP2010} and
Ref.~\onlinecite{Weber-PRB2010}. To minimize the ambiguity caused by
this technical difficulty, we used three different methods; 
Method 1, 2, and 3 (see the Supplementary
Information).   For presentation, we took the average of these three
values as the main data, and the error bars represent the largest and
smallest values obtained by Methods 1--3 in Figure 2.  Importantly,
our conclusions were the same regardless of which values are
considered. In fact, if we consider the previously-used technique,
Method 1, the $U/t$ difference between the RE$_2$CuO$_4$ and
La$_2$CuO$_4$ is enlarged (see the Supplementary
Information).

Arguably, our calculation is the most direct way to determine the
correlation strengths. For the estimation of correlation strength the
previous theoretical approaches analyzed either the mass
renormalization factor or the optical conductivity
\cite{Weber-NP2010,Weber-PRB2010,Das-PRB2009,Comanac-NP2008} with $U$
as a parameter. In the present study, we directly calculated $U$ from
first-principles without any adjustable parameter (see Methods and  Supplementary Information). Therefore, our
results, which show a smaller $U$ value in electron-doped materials,
can be regarded as direct evidence that materials with the $T'$-type
lattice structure are less correlated.

A characteristic feature that determines the
material dependence of the correlation strength can be represented by
a single parameter.  Figure 3(a) shows
the calculated $U/t$ as a function of the inverse of the apical oxygen
height (1/$h_O$) ({\it i.e.}, the average of the inverse bond distance
between apical oxygen and copper).  As 1/$h_O$ increases, the
increasing trend of $U/t$ from the electron-doped materials,
RE$_2$CuO$_4$, to the hole-doped HgBa$_2$CuO$_4$, and to La$_2$CuO$_4$
is obvious.  For the case of RE$_2$CuO$_4$ with no apical oxygen,
1/$h_O$ can be regarded as zero.  While both (hole-doped)
La$_2$CuO$_4$ and HgBa$_2$CuO$_4$ have well-defined octahedral oxygen
cages around the Cu ions ({\it i.e.}, CuO$_6$), no apical oxygen is
found in RE$_2$CuO$_4$, and CuO$_4$ is formed instead of CuO$_6$ (see
Figure 1, inset).  The absence of two apical oxygen atoms can cause a
significant difference in electronic properties and effectively reduce
the correlation strengths.  This relationship between $U/t$ (or $U$)
and $h_O$ can be used as a good rule of thumb to measure the
correlation strength.

It is noteworthy that the hole-doped family can also have
copper-oxygen layers with no apical oxygen. For example, the
inner-layer of HgBa$_2$Ca$_2$Cu$_3$O$_{8}$ has the same local
structure as RE$_2$CuO$_4$ ({\it i.e.}, no apical oxygens;
CuO$_4$). 
Figures 2 and 3(a) clearly show that
the inner-layer Cu in triple-layered HgBa$_2$Ca$_2$Cu$_3$O$_{8}$ has a
similar value of $U$ and $U/t$ to RE$_2$CuO$_4$.

It is a remarkable new finding that some of the hole-doped cuprates
have correlation strengths comparable to the electron-doped materials.
It raises a question about the simple classification that categorizes
all hole-doped cuprates as Mott insulators. 
As shown in Figures 2 and 3(a),
the calculated $U$ and $U/t$ values of the Hg-cuprates are located in
between those of RE$_2$CuO$_4$ and La$_2$CuO$_4$. Note that the
single-layer HgBa$_2$CuO$_4$ has a well-defined CuO$_6$ local unit as
in La$_2$CuO$_4$, and its correlation strength is noticeably weaker
than that of La$_2$CuO$_4$. According to our calculations, the
difference of $U$ ($U/t$) between HgBa$_2$CuO$_4$ and La$_2$CuO$_4$ 
is 1.0 eV (2.1). That difference is  larger  than the 
difference between HgBa$_2$CuO$_4$ and RE$_2$CuO$_4$, which is
$\sim$0.9 eV ($\sim$1.7).  In the case of the triple-layer Hg-compounds,
the correlation strengths decrease to be even closer to the
values of electron-doped materials.  We emphasize its significant
implication for the pairing principle: Considering that the Hg-based
cuprates exhibit quite high $T_c \geq 100 K$, the correlation effects
strong enough to produce the Mott insulating mother compound 
 may not be a prerequisite for high $T_c$ superconductivity.

It is instructive to see how these features are related
to the charge transfer energy, $\Delta_{dp}$ = $E_d$ (Cu-$3d$ energy
level)$-E_p$ (O-$2p$ energy level), which is another key parameter
in many of the transition-metal oxides \cite{Z-S-A}.  While
$\Delta_{dp}$ is a quantity for the $d$-$p$ model ( not  the
single-band model), one can examine the behavior of
$\Delta_{dp}/t$ in comparison to $U/t$.  Figure 3(b) shows the
calculated $\Delta_{dp}/t$ as a function of 1/$h_O$. We note that the
charge transfer energies of the Hg-compounds are more similar to the
values of RE$_2$CuO$_4$ than those of La$_2$CuO$_4$. The overall
behavior of $U$ and $\Delta_{dp}$ is not quite different nor 
entirely similar. the same when plotted
as a function of 1/$h_O$. The similarity is likely due to that 
a large $\Delta_{dp}$ results in a smaller $d$-$p$ hybridization, making
Wannier orbital more localized. At the same time, 
 the details of the band
structure play some role in determining the correlation strength.

Importantly, the results of both $U$ and $\Delta_{dp}$ indicate that
Hg-compounds are significantly less correlated than La$_2$CuO$_4$, and
their correlation strengths are comparable to those of electron-doped
materials.  Therefore, a simple classification of the parent compounds
in terms of the carrier types  is not
pertinent, and the previous studies that regarded La$_2$CuO$_4$ as a
prototype hole-doped cuprate should be re-interpreted. It may be more
desirable to classify some of the hole-doped materials as Slater-type
insulators.

\section*{Discussion}

Comparison of our result with experiments
is not at all straightforward and any direct quantitative argument may
not be possible. The determination of $U$ based on any experimental data
is eventually to fit onto a certain type of model. Within such an
obvious limitation,  
it may be instructive to see the optical conductivity data
as a possible consistency check. The previous experiments
on the hole-doped materials, for example,
seem basically consistent with our results: 
Charge transfer gap
of La$_2$CuO$_4$ is larger than that of Nd$_2$CuO$_4$, and
 the integrated Drude weight 
of (doped) T'-materials is
larger than La$_2$CuO$_4$. The trend of other materials is also compatible with our 
calculations while the data from the undoped parent compounds is not always available
\cite{Comanac-NP2008, Lucarelli-PRL2003, Onose-PRB2004, Cooper-PRB2004, Hwang-PCM2007, Tokura-PRB1990, Uchida-PRB1991}.

Our results can provide natural explanations for recent
experiments
\cite{Brinkmann-PRL1995,Matsumoto-PC2008,Matsumoto-PRB2009,Matsumoto-PC2009,
  Matsumoto-PC2009-2,Matsumoto-PC2010,Yamamoto-SSP2011,Krockenberger-PRB2012,
  Krockenberger-SR2013,Chanda-PRB2014} in which the phase diagram of
the electron-doped cuprates exhibits monotonically increasing $T_c$
toward zero doping (see Figure 1(b)). 
This
behavior has been observed in the carefully-annealed samples of both
thin film and single crystal forms
\cite{Brinkmann-PRL1995,Matsumoto-PC2008,Matsumoto-PRB2009,Matsumoto-PC2009,
	Matsumoto-PC2009-2,Matsumoto-PC2010,Yamamoto-SSP2011,Krockenberger-PRB2012,
	Krockenberger-SR2013,Chanda-PRB2014}. 
If it is indeed the case, the implication can be profound
and  the electron-doped side of the phase diagram
should be re-drawn (Figure 1(b)).
According to our calculations,
this behavior is a result of the relatively weak correlation in the
electron-doped materials.  In this context, it is instructive to
recall a recent numerical result by variational Monte Carlo
calculations.  Yokoyama {\it et al.} showed in their one-band Hubbard
model study that a small value of $U/t\leq 6$ produces an increasing
$T_c$ region of superconductivity whereas a larger $U/t$ value always
gives the dome-shape \cite{Yokoyama-JPSJ2013}.

The Hg-cuprates are of interest in this regard.  Being a hole-doped
family, their correlation strength is significantly weaker than that of
 La$_2$CuO$_4$ and close to the electron-doped cuprates,
especially in the triple-layer compound. Nevertheless, the dome-like doping
dependence of $T_c$ has been observed in both single-layer
\cite{Yamamoto-PRB2000} and multilayer \cite{Mukuda-JPSJ2012}
Hg-cuprates.  Therefore, the dome-shaped $T_c$ may not necessarily be
a consequence of strong electron correlation. In fact, a mechanism
that can induce the dome-shaped $T_c$ without Mott-ness has recently
been proposed \cite{Ogura-arXiv2015}. In this theory, the intrinsic
electron-hole asymmetry of the hybridized Cu$3d$--O$2p$ electronic
structure plays an essential role.  Regarding the absence or presence
of antiferromagnetic ordering, it is important to note that the low
doping regime ($< 5 \%$ ) has not been experimentally reached for the single-layer
Hg-compound due to the presence of excess oxygen
\cite{Yamamoto-PRB2000}.  Hence, considering the moderate value of
$U/t$ in single-layer Hg-cuprates, the presence of antiferromagnetism
as well as the Mott-insulating state in the non-doping limit may still
be an open issue.  We expect the Tl-based cuprate, which also has a
large $h_O$ value, have similar behavior \cite{Shimakawa-PRB1990}.
For multilayer Hg-compounds, antiferromagnetism has been reported in
the underdoped regime \cite{Mukuda-JPSJ2012}.  Our result suggests
that this insulating state can be of the Slater-type rather than the
Mott-type.  The robust presence of antiferromagnetism in these
multilayer cases (compared to the electron-doped cases) might be due
to the interlayer coupling.

\section*{Summary and Conclusion} 
 We performed the first direct calculation of the
material dependent correlation strengths in cuprates.  A clear
increasing trend of $U$ is found as a function of 1/$h_O$. Our result
strongly supports the Slater picture for electron-doped cuprates. It
is the first direct evidence of weaker correlations in electron-doped
materials, and can be regarded as a (theoretical) confirmation. On the
other hand, we significantly revise the current understanding of this
issue. Contrary to the previous conclusion, some of the hole-doped
cuprates ({\it e.g.,} the Hg-compounds) have considerably weaker
correlations which are comparable to those in electron-doped
materials. Our results indicate that the electron correlation strong
enough to induce the Mott gap may not be a prerequisite for high $T_c$
superconductivity.

\section*{Methods} 
\subsection*{Computation details} 
We used so-called `constrained random
phase approximation (cRPA)' method to estimate the correlation
strength. This recently-established technique
\cite{Springer-PRB1998,Kotani-JPCM2000,Aryasetiawan-PRB2004,Aryasetiawan-PRB2006,
  Miyake-PRB2008,Miyake-PRB2009,Sasioglu-PRB2011,Sasioglu-PRB2013,amadon_screened_2014}
has been proven to be reliable in many different types of materials
\cite{Aryasetiawan-PRB2004,Aryasetiawan-PRB2006,
  Miyake-PRB2008,Miyake-PRB2009,Sasioglu-PRB2011,Sasioglu-PRB2013, amadon_screened_2014,
  Vaugier-PRB2012, Sakuma-PRB2013,Werner-PRB2015, Mravlje-PRL2011,
  Martins-PRL2011, Arita-PRL2012, Miyake-JPSJ2008,
  Nakamura-JPSJ2008, Miyake-JPSJ2010, Werner-NP2012}, including 3$d$, 4$d$, 5$d$ transition-metal
oxides \cite{Vaugier-PRB2012,Sakuma-PRB2013,Werner-PRB2015, Mravlje-PRL2011, Martins-PRL2011, Arita-PRL2012} and
Fe-based superconductors \cite{Miyake-JPSJ2008, Nakamura-JPSJ2008,
  Miyake-JPSJ2010,Werner-NP2012}, while it has never been
systematically applied to cuprates. Early
	calculations of La$_2$CuO$_4$ based on constrained LDA (cLDA)
	predict too large $U$ value of $\sim$7--10 eV
 \cite{McMahan-PRB1988, Hybertsen-PRB1989, McMahan-PRB1990, Grant-PRB1992, Anisimov-PRB2002}. It is a typical feature of cLDA due to the limitation
 for describing the electronic screening \cite{Aryasetiawan-PRB2006}. Our implementation of cRPA into our
own software package `{\it
  ecalj}'  \cite{ecalj} follows one of the most recent
standard formalisms by \c{S}a\c{s}{\i}o\u{g}lu {\it et al.}
\cite{Sasioglu-PRB2011,Sasioglu-PRB2013} (see the Supplementary
Information).
We have checked that the previously reported data for many different
materials were well reproduced by our implementation (see the
Supplementary Information).  

In order to avoid the ambiguity related to the $4f$ electrons in RE$_2$CuO$_4$,
we used three different methods. 
Method 1
treats the RE-$4f$ orbitals as the core as in the previous studies
\cite{Weber-NP2010,Weber-PRB2010}. This method removes some screening
channels (but not the on-site $d$-$d$ transitions) around the Fermi
energy and can cause some deviation in the $U$ estimation. Method 2
replaces RE ions with La while maintaining the experimental lattice
parameters. The resulting effect is expected to be similar to Method
1. We emphasize, however, that the whole procedure is determined in a
self-consistent way, and the position and the width of the Cu-$3d$
band is adjusted accordingly. Therefore, the naive guess for the final
$U$ value might not be correct. Method 3 keeps the RE-$4f$ states
around the Fermi energy as described by LDA. Within LDA, these
less-renormalized and uncorrelated $4f$-bands are located closer to
the Fermi level and contribute to the screening. In spite of the
complexity of the LDA band structure, the Cu-$e_g$ bands are well
identified by the standard Wannier fitting, and therefore, Method 3
works as well as the other two approaches 
(see the Supplementary
Information).
The average of these three values is presented as the main data
while the error bars represent the largest and
smallest values obtained by Methods 1--3 (Figure 2).

The LDA band structure was calculated by an
all-electron full-potential method with the PMT basis (augmented plane
wave + muffin-tin orbital) \cite{kotani_pmt_2015}.  The polarization
function is expanded by the mixed product basis in which the imaginary
part along the real axis is accumulated with the tetrahedron method
and the real part is obtained by a Hilbert transformation. Our
approach has a clear advantage in terms of its accuracy compared to
other methods, such as simple $\bfk$-point sampling,
Matsubara-frequency sampling, and the pseudopotential method.  We have
carefully verified the ${\bf k}$-point dependency and found that our
conclusions are robust against the computation details (see the
Supplementary
Information).  The calculated $U$ value of 3.15 eV for
La$_2$CuO$_4$ is in good agreement with the only available data of
3.65 eV \cite{Werner-PRB2015}.  For further details, see the
Supplementary Information.

\section*{Acknowledgments} We thank Dr. Takashi Miyake for providing us
the the maximally localized Wannier function code implemented on top
of `{\it ecalj}' package.  S.W.J. and M.J.H.  were supported by Basic
Science Research Program through the National Research Foundation of
Korea (NRF) funded by the Ministry of Education (2014R1A1A2057202).
The computing resource is supported by National Institute of
Supercomputing and Networking / Korea Institute of Science and
Technology Information with supercomputing resources including
technical support (KSC-2014-C2-015) and by Computing System for
Research in Kyushu University.  T.K. was supported by the Advanced Low
Carbon Technology Research and Development Program (ALCA), the
“High-efficiency Energy Conversion by Spinodal Nano-decomposition”
program of the Japan Science and Technology Agency (JST), and the JSPS
Core-to-Core Program Advanced Research Networks (``Computational
Nano-materials Design on Green Energy'').

\section*{Author contributions}
T.K. and H.K. 
developed the LDA and cRPA code. S.W.J. performed cRPA calculations.
S.W.J. and H.S. calculated $\Delta_{dp}$.  All authors contributed to
analyzing the results and writing the manuscript.

\section*{Additional information}
The authors declare that they have no competing financial interests.

\clearpage

\begin{figure}[h]
\begin{center}
\includegraphics[width=15cm,angle=0]{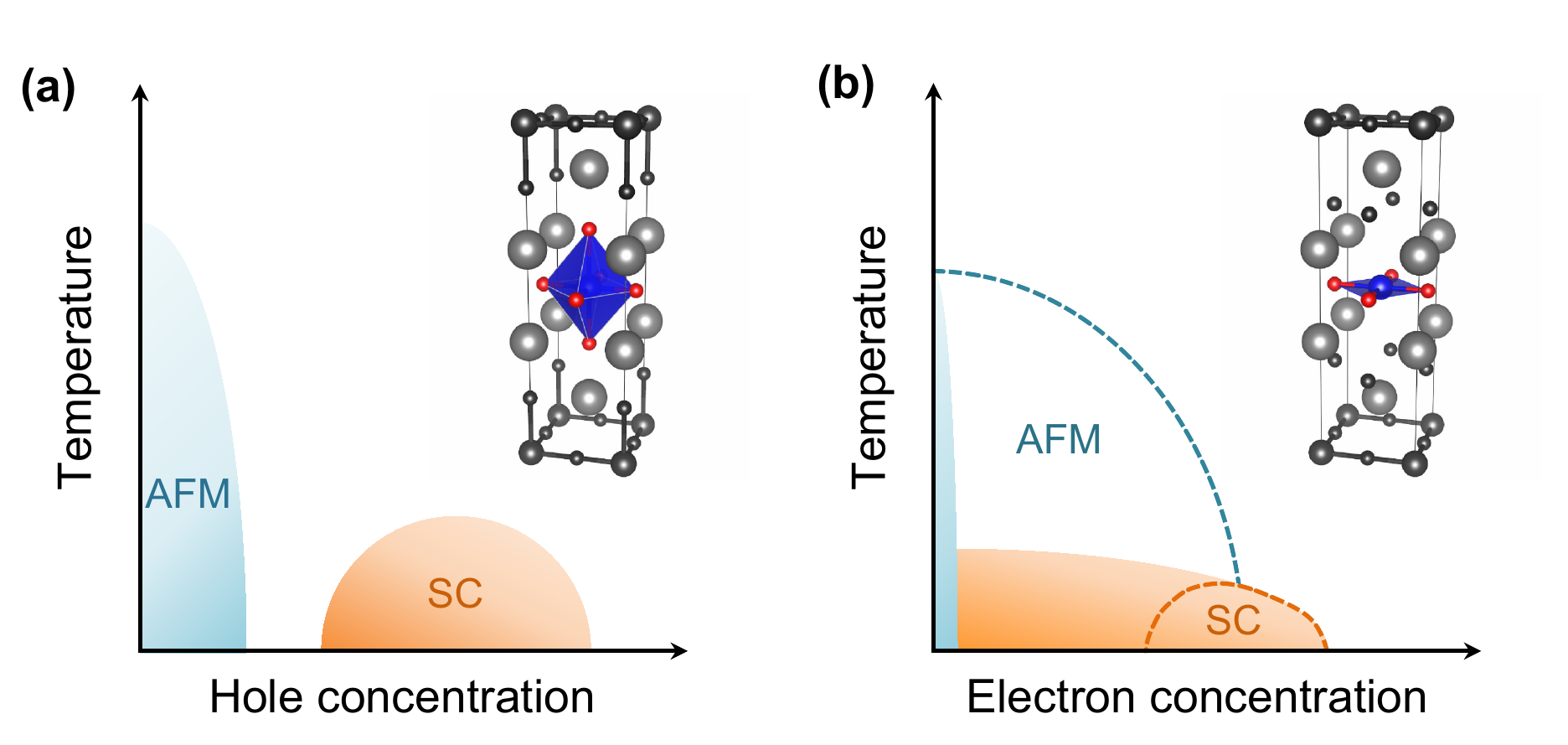}
\caption{Schematic phase diagram of superconducting
  (SC) and antiferromagnetic (AFM) states for the (a) hole-doped and
  (b) electron-doped region. The insets show the representative
  crystal structure for each region: (a) La$_2$CuO$_4$ and (b)
  RE$_2$CuO$_4$ where the large, medium, and small spheres represent
  La/RE (grey), Cu (black or blue), and O (black or red),
  respectively. The octahedral CuO$_6$ and planar CuO$_4$ unit are
  shaded blue.
\label{Figure 1}}
\end{center}
\end{figure}

\begin{figure}[t]
\begin{center}
\includegraphics[width=15cm,angle=0]{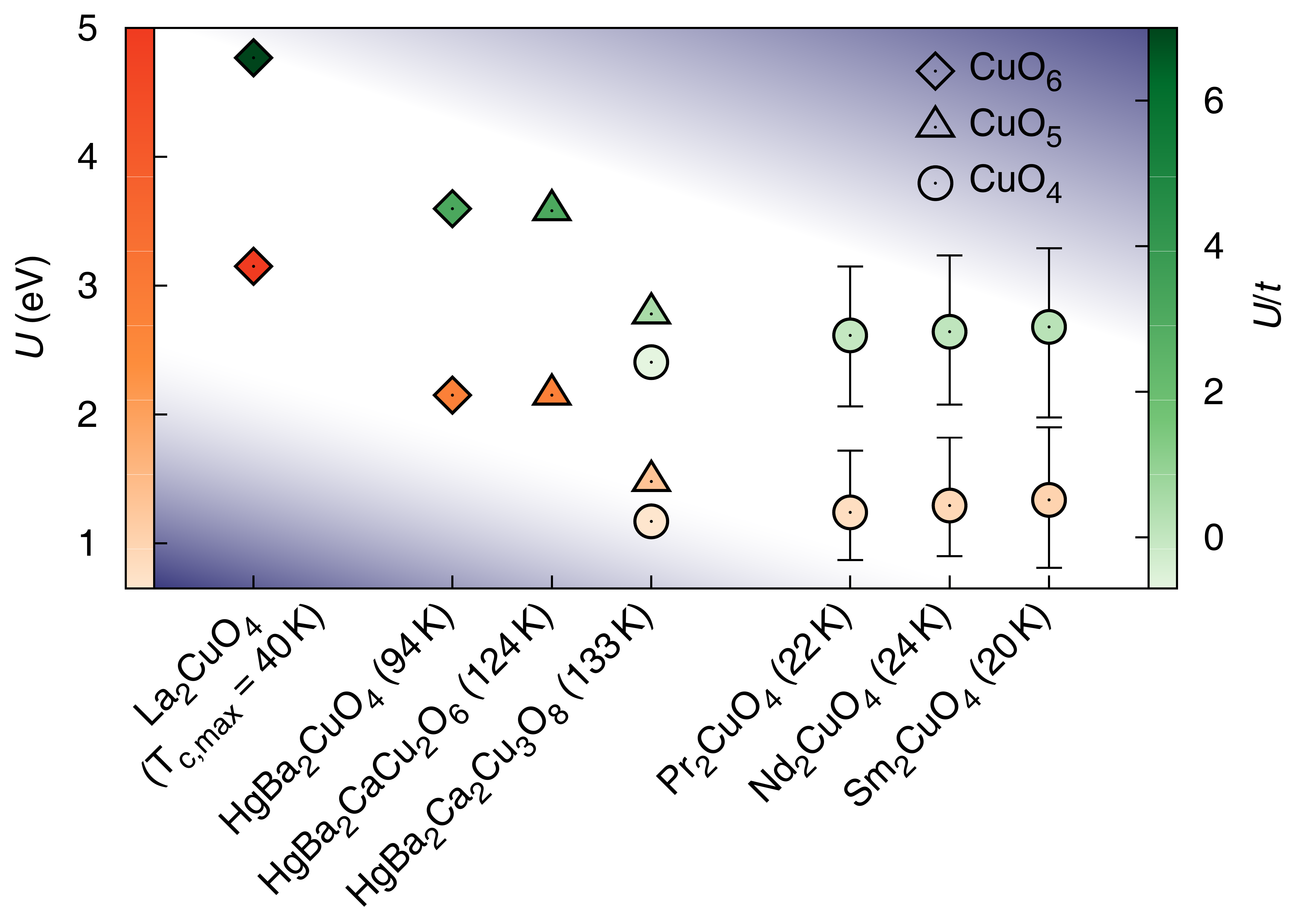}
\caption{  Calculated $U$ and $U$/$t$ for cuprate parent
  compounds. The left (orange) and the right (green) vertical axis
  correspond to $U$ and $U$/$t$, respectively.  A total of seven
  different materials have been calculated: La$_2$CuO$_4$ (single
  layered, hole doped), HgBa$_2$CuO$_4$ (single layered, hole doped),
  HgBa$_2$CaCu$_2$O$_6$ (double layered, hole doped),
  HgBa$_2$Ca$_2$Cu$_3$O$_8$ (triple layered, hole doped),
  Pr$_2$CuO$_4$ (single layered, electron doped), Nd$_2$CuO$_4$
  (single layered, electron doped), and Sm$_2$CuO$_4$ (single layered,
  electron doped). For the electron-doped materials, RE$_2$CuO$_4$,
  three different techniques have been used to treat the RE-$4f$
  electrons (see the text for more details). The average values are
  presented and the error bars indicate the largest and smallest
  values.  The symbols represent the local CuO$_n$ structures:
  diamonds, triangles, and circles correspond to CuO$_6$, CuO$_5$, and
  CuO$_4$, respectively.  The numbers in parentheses are the optimal superconducting $T_{c,\rm max}$ of each material.
\label{Figure 2}}
\end{center}
\end{figure}

\begin{figure}[t]
\begin{center}
\begin{tabular}{c}
  \includegraphics[width=9.5cm,angle=0]{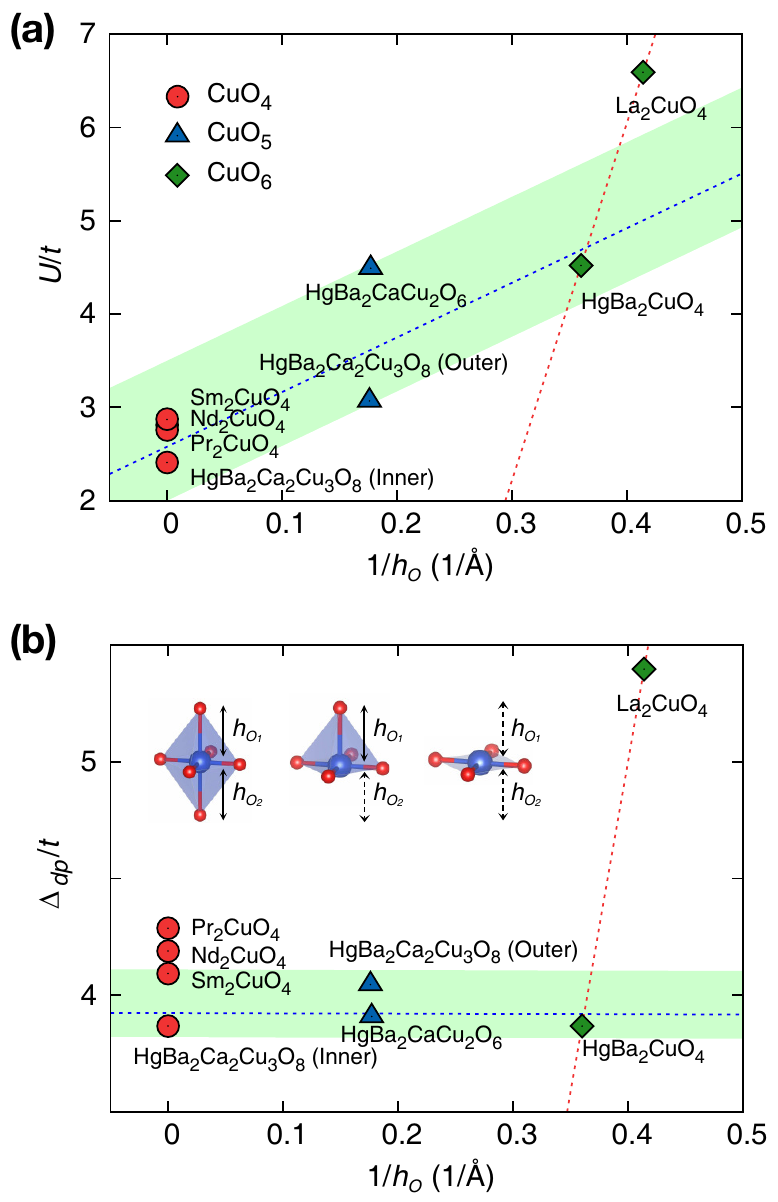}\\
\end{tabular}
\end{center}
    \caption{ \label{Figure 3} The calculated $U/t$ (a) and
      $\Delta_{dp}$/$t$ (b) as a function of the inverse apical oxygen
      height, 1/$h_O$.  The color and shape of each point represent
      the local structure of materials: CuO$_6$ (green diamonds),
      CuO$_5$ (blue triangles), and CuO$_4$ (red circles) having two,
      one, and no apical oxygen, respectively.  The local structures
      are presented in the inset of (b). The effective bond length between Cu and the apical oxygen, $h_O$, is defined as $1/h_O=(1/h_{O_1}+1/h_{O_2})/2$ where $h_{O_{1,2}}$ indicates the Cu to apical oxygen bond distance and the distance can be defined to be $\infty$ when there is no apical oxygen. For the case with no apical
      oxygen (CuO$_4$), 1/$h_O$ can be regarded as zero. For CuO$_5$
      which has one apical oxygen, 1/$h_O$ is defined as half of the
      inverse of the bond distance between Cu and apical O. The red
      line shows the fitting from two data points of single-layer
      hole-doped compounds, La$_2$CuO$_4$ and HgBa$_2$CuO$_4$. The
      blue line shows the fitting from the four data points of the
      Hg-compounds. The shaded green blocks provide a guide for the
      eyes.  }
\end{figure}


\begin{thebibliography}{99}




\bibitem{first1986} Bednorz, J. G. \& M{\"{u}}ller, K. A. Possible high T$_c$ superconductivity in the Ba-La-Cu-O system. {\it Z. Phys. B:
  Condens. Matter} {\bf 64}, 189--193 (1986).



\bibitem{review18} Scalapino, D. J. A common thread: The pairing interaction for unconventional
superconductors. {\it Rev. Mod. Phys.} {\bf 84}, 1383--1417 (2012).


\bibitem{review10} Lee, P. A., Nagaosa, N. \& Wen, X. -G. Doping a Mott insulator: Physics of high-temperature superconductivity.
 {\it Rev. Mod. Phys.} {\bf 78}, 17--85 (2006).

\bibitem{review12} Ogata, M. \& Fukuyama, H. The $t$-$J$ model for the oxide high-T$_c$
superconductors. {\it Rep. Prog. Phys.} {\bf 71}, 036501 (2008).

\bibitem{review17} Armitage, N. P., Fournier, P. \& Greene, R. L. Progress and perspectives on electron-doped cuprates. {\it Rev. Mod. Phys.} {\bf 82}, 2421--2487 (2010).


\bibitem{review-PhysicaC} Fournier, P. T$'$ and infinite-layer electron-doped cuprates.
{\it Physica C} {\bf 514}, 314--338 (2015).





\bibitem{Brinkmann-PRL1995}
Brinkmann, M., Rex, T., Bach, H. \& Westerholt, K.
Extended superconducting concentration range observed in Pr$_{2-x}$Ce$_x$CuO4.
{\it Phys. Rev. Lett.} {\bf 74}, 4927--4930 (1995). 



\bibitem{Matsumoto-PC2008}
Matsumoto, O. {\it et al.}
Superconductivity in undoped $T'$-$RE_2$CuO$_4$ with $T_c$ over 30K.
{\it Physica C} {\bf 468}, 1148--1151 (2008). 

\bibitem{Matsumoto-PRB2009}
Matsumoto, O. {\it et al.}
Synthesis and properties of superconducting $T'$-$R_2$CuO$_4$ ($R$=Pr, Nd, Sm, Eu, Gd).
{\it Phys. Rev. B} {\bf 79}, 100508(R) (2009).

\bibitem{Matsumoto-PC2009}
Matsumoto, O. {\it et al.}
Generic phase diagram of ``electron-doped'' T$'$ cuprates.
{\it Physica C} {\bf 469}, 924--927 (2009).

\bibitem{Matsumoto-PC2009-2}
Matsumoto, O. {\it et al.}
Reduction dependence of superconductivity in the end-member T$'$ cuprates.
{\it Physica C} {\bf 469}, 940--943 (2009).

\bibitem{Matsumoto-PC2010}
Matsumoto, O., Tsukada, A., Yamamoto, H., Manabe, T. \& Naito, M.
Generic phase diagram of Nd$_{2-x}$Ce$_x$CuO$_4$.
{\it Physica C} {\bf 470}, S101--S103 (2010).

\bibitem{Yamamoto-SSP2011}
Yamamoto, H., Matsumoto, O., Krockenberger, Y., Yamagami, K. \& Naito, M.
Molecular beam epitaxy of superconducting Pr$_2$CuO$_4$ films.
{\it Solid State Commun.} {\bf 151}, 771–774 (2011).

\bibitem{Krockenberger-PRB2012}
Krockenberger, Y., Yamamoto, H., Tsukada, A., Mitsuhashi, M. \& Naito, M.
Unconventional transport and superconducting properties in electron-doped cuprates.
{\it Phys. Rev. B} {\bf 85}, 184502 (2012).

\bibitem{Krockenberger-SR2013}
Krockenberger, Y. {\it et al.}
Emerging superconductivity hidden beneath charge-transfer insulators.
{\it Sci. Rep.} {\bf 3}, 2235 (2013).

\bibitem{Chanda-PRB2014}
Chanda, G. {\it et al.}
Optical study of superconducting Pr$_2$CuO$_x$ with $x \simeq 4$.
{\it Phys. Rev. B} {\bf 90}, 024503 (2014).


\bibitem{Tsukada-SSC2005}
 Tsukada, A. {\it et al.}
 New class of T$'$-structure cuprate superconductors.
 {\it Solid State Commun.} {\bf 133}, 427--431 (2005).

\bibitem{Adachi-JPSJ2013}
 Adachi, T. {\it et al.}
 Evolution of the electronic state through the reduction annealing in electron-doped Pr$_{1.3-x}$La$_{0.7}$Ce$_x$CuO$_{4+\delta}$ ($x =$ 0.10) Single Crystals: Antiferromagnetism, Kondo Effect, and Superconductivity.
 {\it J. Phys. Soc. Jpn.} {\bf 82}, 063713 (2013).



\bibitem{Weber-NP2010}
Weber, C., Haule, K. \& Kotliar, G.
Strength of correlations in electron- and hole-doped cuprates.
{\it Nature Phys.} {\bf 6}, 574--578 (2010).

\bibitem{Weber-PRB2010}
Weber, C., Haule, K. \& Kotliar, G.
Apical oxygens and correlation strength in electron- and hole-doped copper oxides.
{\it Phys. Rev. B} {\bf 82}, 125107 (2010).


\bibitem{Das-PRB2009}
Das, H. \& Saha-Dasgupta, T.
Electronic structure of La$_2$CuO$_4$ in the $T$ and $T'$ crystal structures using dynamical mean field theory.
{\it Phys. Rev. B} {\bf 79}, 134522 (2009).


\bibitem{Comanac-NP2008}
Comanac, A., de' Medici, L., Capone, M. \& Millis, A. J. 
Optical conductivity and the correlation strength of high-temperature copper-oxide superconductors.
{\it Nature Phys.} {\bf 4}, 287--290 (2008).






\bibitem{Marzari-PRB1997}
Marzari, N. \& Vanderbilt, D.
Maximally localized generalized Wannier functions for composite energy bands.
{\it Phys. Rev. B} {\bf 56}, 12847--12865 (1997).

\bibitem{Souza-PRB2001}
Souza, I., Marzari, N. \& Vanderbilt, D.
Maximally localized Wannier functions for entangled energy bands.
{\it Phys. Rev. B} {\bf 65}, 035109 (2001).



\bibitem{Araujo-EPL2012}
Ara{\' u}jo, M. A. N., Carmelo, J. M. P., Sampaio, M. J. \& White, S. R.
Spin-spectral-weight distribution and energy range of the parent compound La$_2$CuO$_4$.
{\it  Eur. Phys. Lett.} {\bf 98}, 67004 (2012).


\bibitem{Z-S-A} 
Zaanen, J., Sawatzky, G. A.  \& Allen, J. W., 
Band gaps and electronic structure of transition-metal compounds.
{\it Phys. Rev. Lett.} {\bf 55}, 418 (1985).




\bibitem{Lucarelli-PRL2003}
Lucarelli, A. {\it et al.}
Phase diagram of La$_{2-x}$Sr$_x$CuO$_4$ probed in the infrared: Imprints of charge stripe excitations.
{\it Phys. Rev. Lett.} {\bf 90}, 037002 (2003).
\bibitem{Onose-PRB2004}
Onose, Y., Taguchi, Y., Ishizaka, K. \& Tokura, Y.
Charge dynamics in underdoped Nd$_{2−x}$Ce$_x$CuO$_4$: Pseudogap and related phenomena. 
{\it Phys. Rev. B} {\bf 69}, 024504 (2004).
\bibitem{Cooper-PRB2004}
Cooper, S. L. {\it et al.}
Optical studies of the a-, b-, and c-axis charge dynamics in YBa$_2$Cu$_3$O$_{6+x}$.
{\it Phys. Rev. B} {\bf 47}, 8233--8248 (1993).
\bibitem{Hwang-PCM2007}
Hwang, J., Timusk, T. \& Gu, G. D. J.
Doping dependent optical properties of Bi$_2$Sr$_2$CaCu$_2$O$_{8+\delta}$.
{\it Phys. Condens. Matter} 19, 125208 (2007).
\bibitem{Tokura-PRB1990}
Tokura, Y. {\it et al.} 
Cu-O network dependence of optical charge-transfer gaps and spin-pair excitations in single-CuO$_2$-layer compounds.
{\it Phys. Rev. B} {\bf 41}, 11657(R) (1990).
\bibitem{Uchida-PRB1991}
Uchida, S. {\it et al.} 
Optical spectra of La$_{2-x}$Sr$_x$CuO$_4$: Effect of carrier doping on the electronic structure of the CuO$_2$ plane.
{\it Phys. Rev. B} {\bf 43}, 7942 (1991).



\bibitem{Yokoyama-JPSJ2013}
Yokoyama, H., Ogata, M., Tanaka, Y., Kobayashi, K. \& Tsuchiura, H.
Crossover between BCS Superconductor and Doped Mott Insulator of $d$-Wave Pairing State in Two-Dimensional Hubbard Model.
{\it J. Phys. Soc. Jpn.} {\bf 82}, 014707 (2013).


\bibitem{Yamamoto-PRB2000}
Yamamoto A., Hu, W. -Z. \&  Tajima, S. 
Thermoelectric power and resistivity of HgBa$_2$CuO$_{4+\delta}$ over a wide doping range. {\it Phys. Rev. B} {\bf 63}, 024504 (2000).


\bibitem{Mukuda-JPSJ2012}
Mukuda, H., Shimizu, S., Iyo, A. \& Kitaoka, Y.
High-$T_c$ superconductivity and antiferromagnetism in multilayered copper oxides --A new paradigm of superconducting mechanism--.
{\it J. Phys. Soc. Jpn.} {\bf 81}, 011008 (2012).




\bibitem{Ogura-arXiv2015}
Ogura, D. \& Kuroki, K.
Asymmetry of superconductivity in hole- and electron-doped cuprates : explanation within two-particle self-consistent analsys for the three band model. {\it arXiv}: 1505.04017.


\bibitem{Shimakawa-PRB1990}
Shimakawa, Y., Kubo, Y., Manako, T., \& Igarashi, H., 
Variation in $T_{C}$ and carrier concentration in Tl based sperconductors.
{\it Phys. Rev. B} {\bf 40}, 11400(R) (1989).





\bibitem{Springer-PRB1998}
Springer, M. \& Aryasetiawan, F.
Frequency-dependent screened interaction in Ni within the random-phase approximation.
{\it Phys. Rev. B} {\bf 57}, 4364--4368 (1998). 



\bibitem{Kotani-JPCM2000}
Kotani, T.
{\it Ab initio} random-phase-approximation calculation of the frequency-dependent effective interaction between 3d electrons: Ni, Fe, and MnO.
{\it J. Phys.: Condens. Matter} {\bf 12}, 2413--2422 (2000).



\bibitem{Aryasetiawan-PRB2004}
Aryasetiawan, F. {\it et al.}
Frequency-dependent local interactions and low-energy effective models from electronic structure calculations.
{\it Phys. Rev. B} {\bf 70}, 195104 (2004). 

\bibitem{Aryasetiawan-PRB2006}
Aryasetiawan, F., Karlsson, K., Jepsen, O. \& Sch{\"{o}}nberger, U.
Calculations of Hubbard $U$ from first-principles.
{\it Phys. Rev. B} {\bf 74}, 125106 (2006).



\bibitem{Miyake-PRB2008}
Miyake, T. \& Aryasetiawan, F.
Screened Coulomb interaction in the maximally localized Wannier basis.
{\it Phys. Rev. B} {\bf 77}, 085122 (2008).

\bibitem{Miyake-PRB2009}
Miyake, T., Aryasetiawan, F. \& Imada, M.
{\it Ab initio} procedure for constructing effective models of correlated materials with entangled band structure.
{\it Phys. Rev. B} {\bf 80}, 155134 (2009).

\bibitem{Sasioglu-PRB2011}
\c{S}a\c{s}{\i}o\u{g}lu, E., Friedrich, C. \& Bl{\"{u}}gel, S.
Effective Coulomb interaction in transition metals from constrained random-phase approximation.
{\it Phys. Rev. B} {\bf 83}, 121101(R) (2011).


\bibitem{Sasioglu-PRB2013}
\c{S}a\c{s}{\i}o\u{g}lu, E., Galanakis, I., Friedrich, C. \& Bl{\"{u}}gel, S.
{\it Ab initio} calculation of the effective on-site Coulomb interaction parameters for half-metallic magnets.
{\it Phys. Rev. B} {\bf 88}, 134402 (2013).




\bibitem{amadon_screened_2014}
 Amadon, B., Applencourt, T. \& Bruneval, F.
 Screened {Coulomb} interaction calculations: {cRPA} implementation and
 applications to dynamical screening and self-consistency in uranium dioxide and cerium.
 {\it Phys. Rev. B} {\bf 89}, 125110 (2014).


\bibitem{Vaugier-PRB2012}
Vaugier, L., Jiang, H. \& Biermann, S.
Hubbard $U$ and Hund exchange $J$ in transition metal oxides: Screening versus localization trends from constrained random phase approximation.
{\it Phys. Rev. B} {\bf 86}, 165105 (2012).


\bibitem{Sakuma-PRB2013}
Sakuma, R. \& Aryasetiawan, F.
First-principles calculations of dynamical screened interactions for the transition metal oxides $M$O ($M$=Mn, Fe, Co, Ni).
{\it Phys. Rev. B} {\bf 87}, 165118 (2013).



\bibitem{Werner-PRB2015}
Werner, P., Sakuma, R., Nilsson, F. \&  Aryasetiawan, F.
Dynamical screening in La$_2$CuO$_4$
{\it Phys. Rev. B} {\bf 91}, 125142 (2015).



\bibitem{Mravlje-PRL2011}
Mravlje, J. {\it et al.}
Coherence-incoherence crossover and the mass-renormalization puzzles in Sr$_2$RuO$_4$.
{\it Phys. Rev. Lett.} {\bf 106}, 096401 (2011).

\bibitem{Martins-PRL2011}
Martins, C., Aichhorn, M., Vaugier, L. \& Biermann, S.
Reduced effective spin-orbital degeneracy and spin-orbital ordering in paramagnetic transition-metal oxides: Sr$_2$IrO$_4$ versus Sr$_2$RhO$_4$.
{\it Phys. Rev. Lett.} {\bf 107}, 266404 (2011).

\bibitem{Arita-PRL2012}
Arita, R., Kune\v{s}, J., Kozhevnikov, V., Aichhorn, M., Eguiluz, A. G. \& Imada, M.
{\it Ab initio} studies on the interplay between spin-orbit interaction and Coulomb correlation in Sr$_2$IrO$_4$ and Ba$_2$IrO$_4$.
{\it Phys. Rev. Lett.} {\bf 108}, 086403 (2012).






\bibitem{Miyake-JPSJ2008}
Miyake, T., Pourovskii, L., Vildosola, V., Biermann, S. \& Georges, A.
d- and f-orbital correlations in the REFeAsO compounds.
{\it J. Phys. Soc. Jpn.} {\bf 77}, 99--102 (2008).

\bibitem{Nakamura-JPSJ2008}
Nakamura, K., Arita, R. \& Imada, M.
{\it Ab initio} derivation of low-energy model for iron-based superconductors LaFeAsO and LaFePO.
{\it J. Phys. Soc. Jpn.} {\bf 77}, 093711 (2008).


\bibitem{Miyake-JPSJ2010}
Miyake, T., Nakamura, K., Arita, R. \& Imada, M.
Comparison of {\it Ab initio} low-energy models for LaFePO, LaFeAsO, BaFe$_2$As$_2$, LiFeAs, FeSe, and FeTe: electron correlation and covalency.
{\it J. Phys. Soc. Jpn.} {\bf 79}, 044705 (2010).
     

\bibitem{Werner-NP2012}
Werner, P. {\it et al.}
Satellites and large doping and temperature dependence of electronic properties in hole-doped BaFe$_2$As$_2$.
{\it Nature Phys.} {\bf 8}, 331--337 (2012).




\bibitem{McMahan-PRB1988}
McMahan, A. K., Martin, R. M. \& Satpathy, S.
Calculated effective Hamiltonian for La$_2$CuO$_4$ and solution in the impurity Anderson approximation.
{\it Phys. Rev. B} {\bf 38}, 6650 (1988).

\bibitem{Hybertsen-PRB1989}
Hybertsen, M. S., Schl{\"u}ter, M. \& Christensen, N. E.
Calculation of Coulomb-interaction parameters for La$_2$CuO$_4$ using a constrained-density-functional approach.
{\it Phys. Rev. B} {\bf 39}, 9028 (1989).

\bibitem{McMahan-PRB1990}
McMahan, A. K., Annett, J. F. \& Martin, R. M.
Cuprate parameters from numerical Wannier functions.
{\it Phys. Rev. B} {\bf 42}, 6268 (1990).

\bibitem{Grant-PRB1992}
Grant, J. B. \& McMahan, A. K.
Spin bags and quasiparticles in doped La$_2$CuO$_4$.
{\it Phys. Rev. B} {\bf 46}, 8440 (1992).

\bibitem{Anisimov-PRB2002}
Anisimov, V. I., Korotin, M. A., Nekrasov, I. A., Pchelkina, Z. V. \& Sorella, S.
First principles electronic model for high-temperature superconductivity.
{\it Phys. Rev. B} {\bf 66}, 100502(R) (1990).



\bibitem{ecalj}
Kotani, T., {\it ecalj} package. Available at: https://github.com/tkotani/ecalj (2009).




\bibitem{kotani_pmt_2015}
  Kotani, T., Kino, H. \& Akai, H.
  Formulation of the augmented plane-wave and muffin-tin orbital method.
  {\it J. Phys. Soc. Jpn.} {\bf 84}, 034702 (2015).





\end{thebibliography}
\end{document}


\title{Supplementary information:\\Direct theoretical evidence for weaker correlations in
  electron-doped and Hg-based hole-doped cuprates}

\author{Seung Woo Jang}
\affiliation{Department of Physics, Korea Advanced Institute of
  Science and Technology (KAIST), Daejeon 305-701, Korea }

\author{Hirofumi Sakakibara} \affiliation{Computational Condensed
  Matter Physics Laboratory, RIKEN, Wako, Saitama 351-0198, Japan
}

\author{Hiori Kino} \affiliation{National Institute for Materials
  Science, Sengen 1-2-1, Tsukuba, Ibaraki 305-0047, Japan.}

\author{Takao Kotani} \affiliation{Department of Applied Mathematics
  and Physics, Tottori University, Tottori 680-8552, Japan}

\author{Kazuhiko Kuroki} 
\affiliation{Department of Physics, Osaka University,
  Machikaneyama-Cho, Toyonaka, Osaka 560-0043, Japan}

\author{Myung Joon Han} \email{mj.han@kaist.ac.kr}
\affiliation{Department of Physics, Korea Advanced Institute of
  Science and Technology (KAIST), Daejeon 305-701, Korea }
\affiliation{ KAIST Institute for the NanoCentury, Korea Advanced
  Institute of Science and Technology, Daejeon 305-701, Korea }



\maketitle

\begin{figure}[t]
\begin{center}
\includegraphics[width=16cm,angle=0]{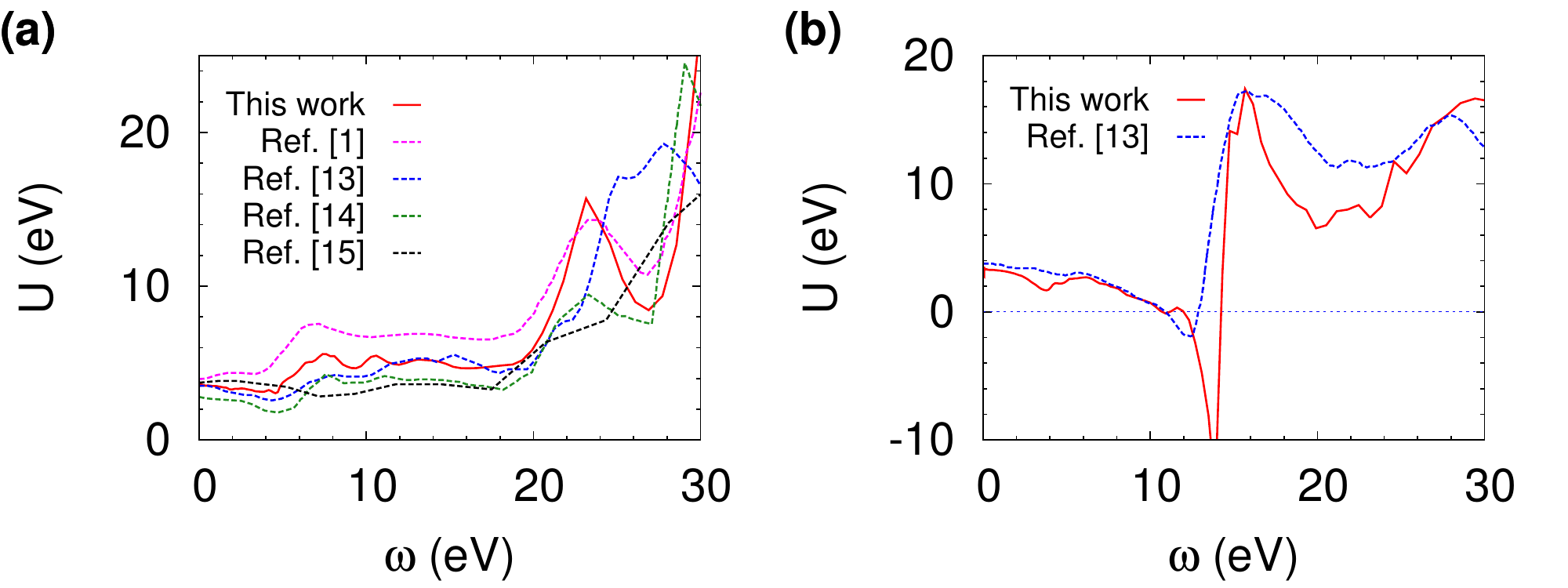}
\renewcommand{\thefigure}{S\arabic{figure}}
\caption{ (Color online) The calculated $U(\omega)$ (red solid lines) for (a)
  paramagnetic Ni and (b) SrVO$_3$ in comparison with the previous
  studies (dashed lines). The five $3d$ states and the three $t_{2g}$
  states are averaged for (a) Ni and (b) SrVO$_3$, respectively.}
\label{fig:Reproduce}
\end{center}
\end{figure}

\begin{table}[t]
\renewcommand{\thetable}{S\arabic{table}}
\caption{The calculated values of $U(\omega=0)$ for paramagnetic Ni
  and SrVO$_3$}
\vspace{0.3cm}
\begin{ruledtabular}
\begin{tabular}{ccccccc}
 						& \multicolumn{6}{c}{$U$ (eV)} \\
 						\cline{2-7}
Material	 & This work & Ref.~\cite{Aryasetiawan-PRB2006} & Ref.~\cite{Miyake-PRB2008} & Ref.~\cite{Miyake-PRB2009} & Ref.~\cite{Sasioglu-PRB2011} & Ref.~\cite{Vaugier-PRB2012}  \\\hline
	
		 Ni		& 3.56 & 3.7 & 2.77 & 3.40--4.05\footnote{Miyake {\it et al.}~\cite{Miyake-PRB2009} showed that the static $U$ can be slightly varied with  the different energy windows chosen for Wannier functions.} & 3.96 & N/A \\ 
	     SrVO$_3$	& 3.15 & 3.5 & 3.0  & N/A  & N/A & 3.2 \\
\end{tabular}
\end{ruledtabular}
\label{tab:Reproduce}
\end{table}

\section{Computation details}

We implemented a recent cRPA formalism refined by
\c{S}a\c{s}{\i}o\u{g}lu {\it et al.}
\cite{Sasioglu-PRB2011,Sasioglu-PRL2012,Sasioglu-PRB2013} in a
first-principles electronic structure calculation package `{\tt ecalj}'
\cite{ecalj-1,ecalj-2}, which is originally designed for 
quasiparticle self-consistent GW calculations \cite{kotani_qsgw_2014}.
Our method has computational advantage by keeping the positive
definiteness of the imaginary part of screened Coulomb interaction; 
the same technique with that of Ref.\cite{Kotani-JPCM2000}.  In all
calculations, we used the experimental crystal structures
\cite{Jorgensen-PRL1987, Wagner-PC1993, Cantoni-PC1993,
  Kaluzhskikh-JSSC2011, Chattopadhyay-PC1991}. The interaction
parameter, $U_{x^2-y^2}$, is estimated by both one band fitting and
two band ($e_g$) fitting, and the results are in good agreement with
each other. The most time-consuming check procedure is about the ${\bf
  k}$-point convergence.  We have carefully performed this check (see
Section II in the below). The number of ${\bf k}$ points used in 
cRPA calculations are 8$\times$8$\times$8, 8$\times$8$\times$4,
8$\times$8$\times$3, 8$\times$8$\times$2, 8$\times$8$\times$8 for the first Brillouin
zone of La$_2$CuO$_4$, HgBa$_2$CuO$_4$, HgBa$_2$CaCu$_2$O$_6$,
HgBa$_2$Ca$_2$Cu$_3$O$_8$, and RE$_2$CuO$_4$ respectively. For LDA calculations, the number of ${\bf k}$ points were increased;
12$\times$12$\times$12, 12$\times$12$\times$8,
12$\times$12$\times$4, 12$\times$12$\times$4, and
12$\times$12$\times$12 for La$_2$CuO$_4$, HgBa$_2$CuO$_4$,
HgBa$_2$CaCu$_2$O$_6$, HgBa$_2$Ca$_2$Cu$_3$O$_8$, and RE$_2$CuO$_4$,
respectively. Since the ${\bf k}$-points for the multilayer
Hg-compounds could not be as many as for the others due to the
large computational cost, 
our results for these cases could be slightly overestimated (see
Section II). However, it will not affect any of our conclusions as
clearly seen in Section II.
Our code has been tested with paramagnetic Ni and SrVO$_3$ which
are the most
extensively-examined systems in the literature
\cite{Aryasetiawan-PRB2006,Miyake-PRB2008,Miyake-PRB2009,
  Sasioglu-PRB2011,Vaugier-PRB2012}. As shown in
Table \ref{tab:Reproduce} and Fig.~\ref{fig:Reproduce}, our results are
in good agreement with the previous ones.

\begin{figure}[!ht]
\begin{center}
\includegraphics[width=11cm,angle=0]{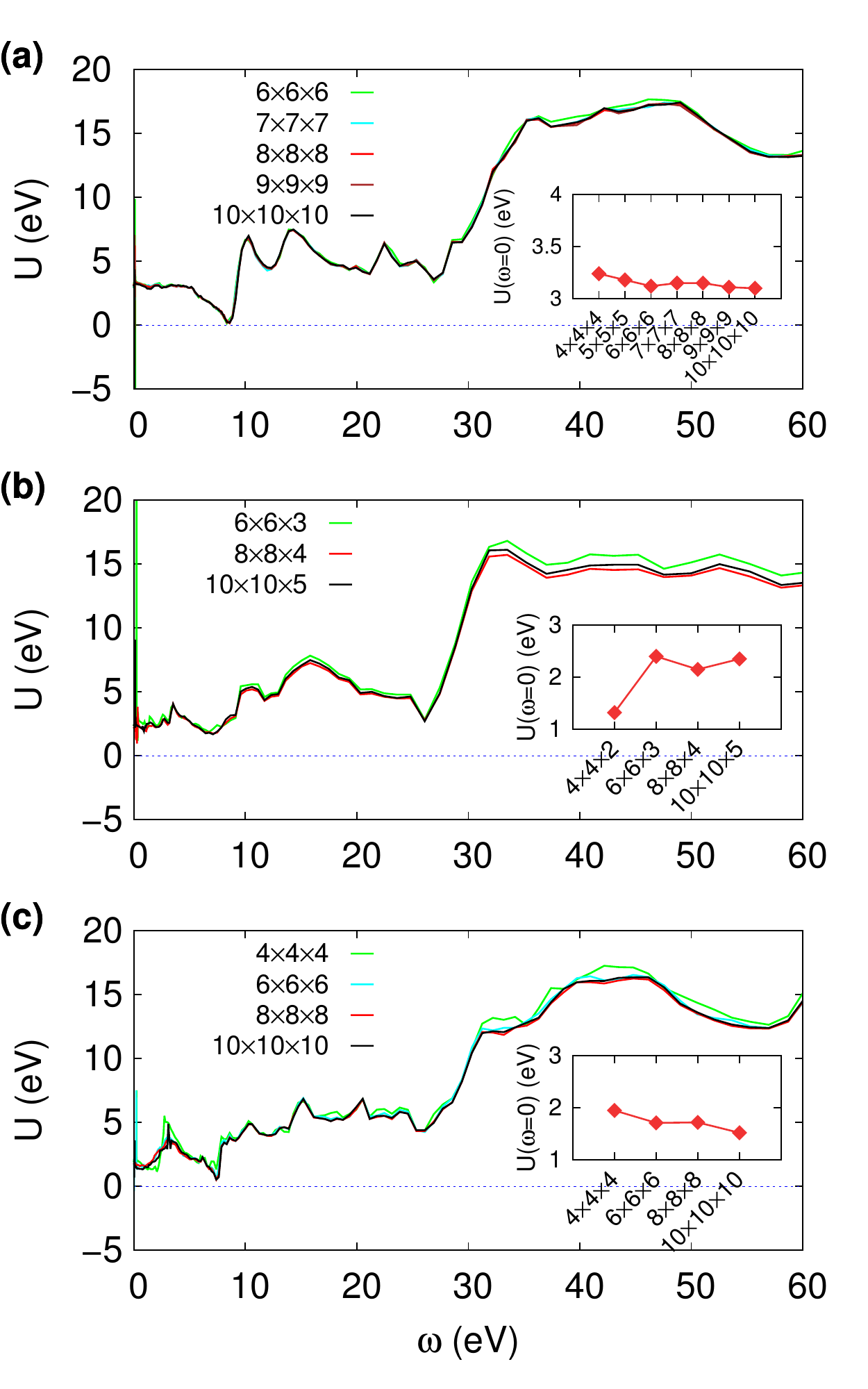}
\renewcommand{\thefigure}{S\arabic{figure}}
\caption{ (Color online) The calculated $U(\omega)$ with different numbers of ${\bf
    k}$ points for (a) La$_2$CuO$_4$, (b) HgBa$_2$CuO$_4$, and (c)
  Pr$_2$CuO$_4$ (Method 3). The insets show the calculated
  $U(\omega=0)$ corresponding to each ${\bf k}$ mesh.}
\label{fig:Kpoint}
\end{center}
\end{figure}

\section{k-point test}
We have 
carefully checked the ${\bf k}$-point dependence of $U(\omega)$ for
all materials considered in this study. The systematic
behaviors are always found as the number of ${\bf k}$-points 
increases as shown in Fig.~\ref{fig:Kpoint} where we chose
La$_2$CuO$_4$ (a), HgBa$_2$CuO$_4$ (b) and Pr$_2$CuO$_4$ (Method 3) (c)
as the representative examples. This test calculation clearly shows
that the ${\bf k}$ meshes we used is good enough with only two
exceptions of HgBa$_2$CaCu$_2$O$_{6}$ and
HgBa$_2$Ca$_2$Cu$_3$O$_{8}$. As discussed in the main text, the ${\bf
  k}$ meshes used for these multilayer Hg-cuprates may not be enough
and the $U$ values get slightly reduced by increasing the ${\bf k}$
points. However our ${\bf k}$-point test shows that the change would
not be significant (presumably $\sim$0.1 eV) and our conclusions not
be changed.

\begin{figure}[!t]
\begin{center}
\renewcommand{\thefigure}{S\arabic{figure}}
\includegraphics[width=12cm,angle=0]{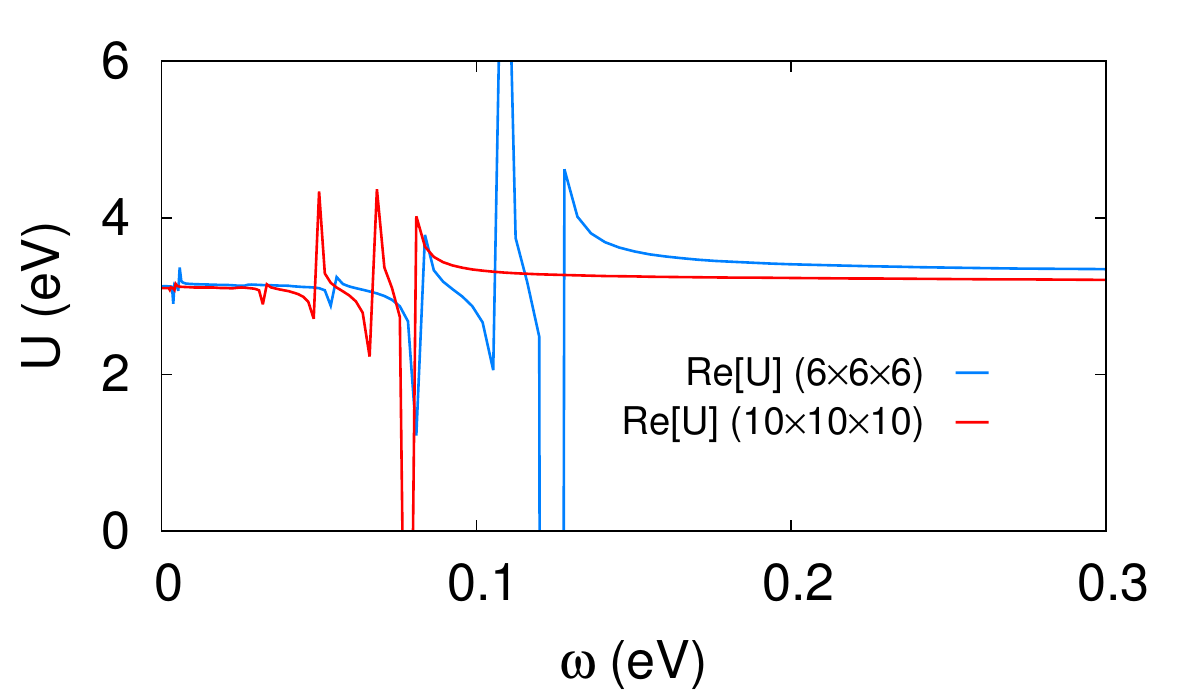}
\caption{The calculated $U(\omega)$ of La$_2$CuO$_4$ in the low
frequency regime. The blue and red line
represents the results from 6$\times$6$\times$6 and 10$\times$10$\times$10 $\bf k$-mesh, respectively. The oscillation due to the remnantal screening becomes less significant as the number of $\bf k$ points increases and at the lower frequency. The error inevitably caused by this kind of oscillation is expected to be $\sim$0.1 eV (see text for more details).}
\label{fig:Low_frequency}
\end{center}
\end{figure}

 We paid special attention to the low frequency oscillation of $U(\omega)$ (Fig.\ref{fig:Low_frequency}). They are observed in our implemnetation most likely due to the incomplete matching between the original $d$ and Wannier bands
 \cite{Sasioglu-PRB2011, Sasioglu-PRL2012, Sasioglu-PRB2013} although such low energy excitations should in principle
be absent within the original spirit of cRPA. The size and the position of remnant excitations
are related to the discrete $\bf k$ sum around Fermi surface, and they become less pronounced as
we use larger number of $\bf k$ points (Fig.\ref{fig:Low_frequency}). Also from the other check calculations based on
Kramers-Kronig relation, we conclude that the error caused by this excitation is just about $\sim$0.1 eV.

\section{Wannier fitting}
\begin{figure}[!t]
\begin{center}
\renewcommand{\thefigure}{S\arabic{figure}}
\includegraphics[width=12cm,angle=0]{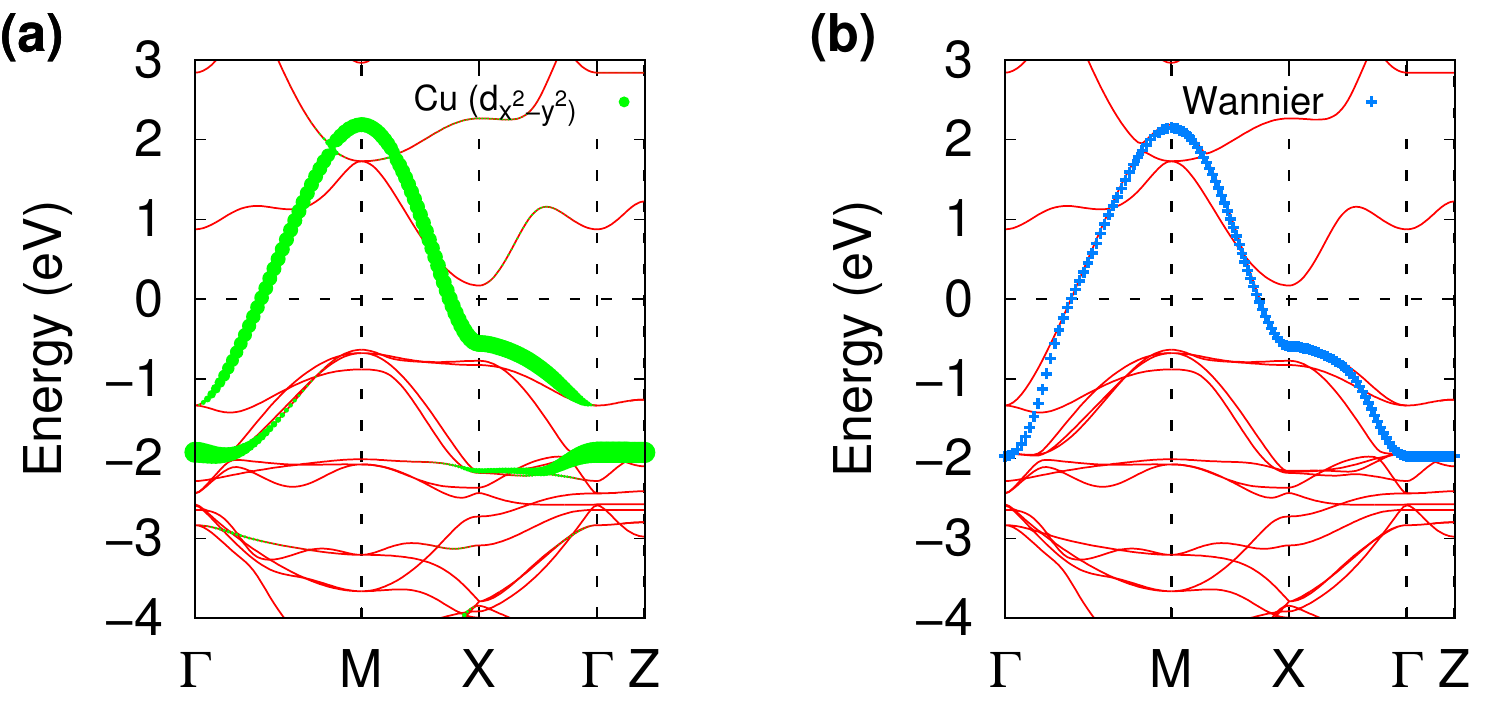}
\includegraphics[width=12cm,angle=0]{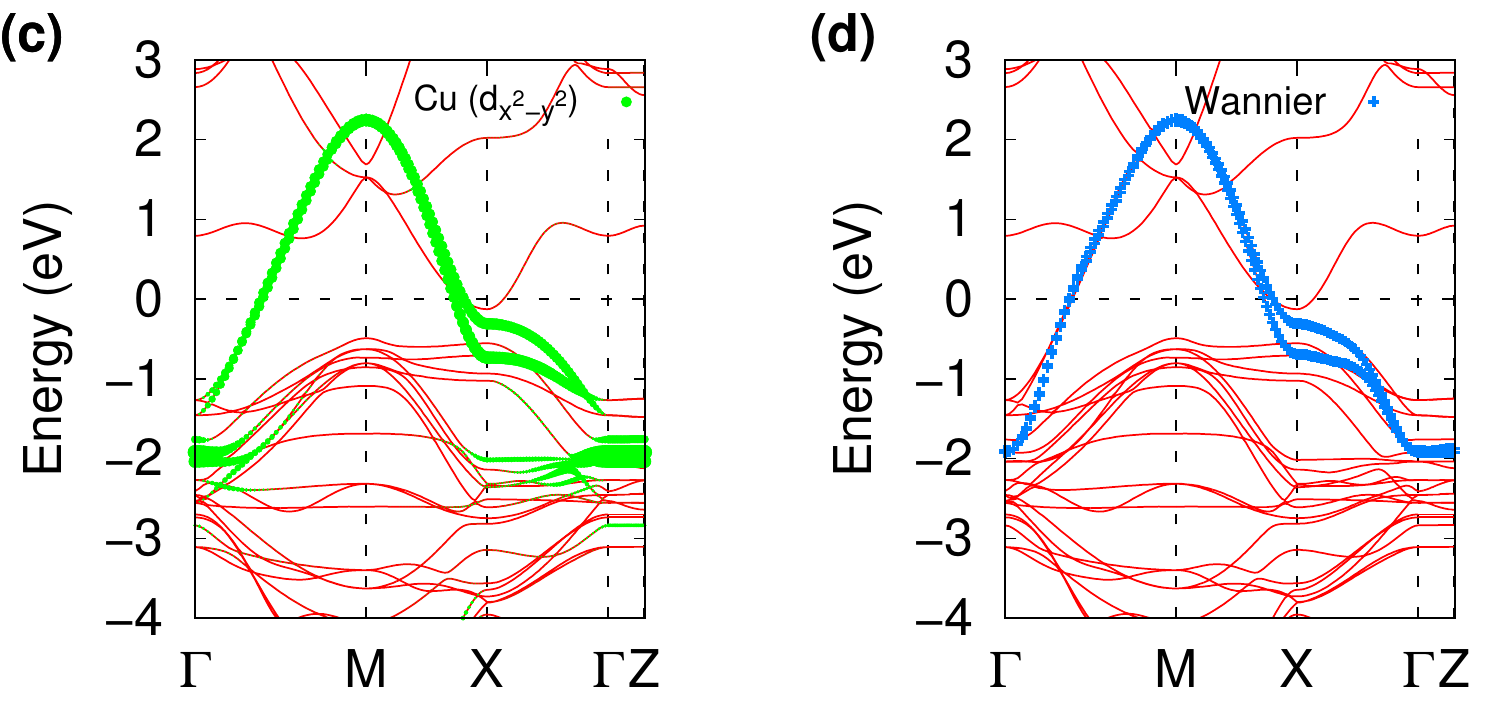}
\includegraphics[width=12cm,angle=0]{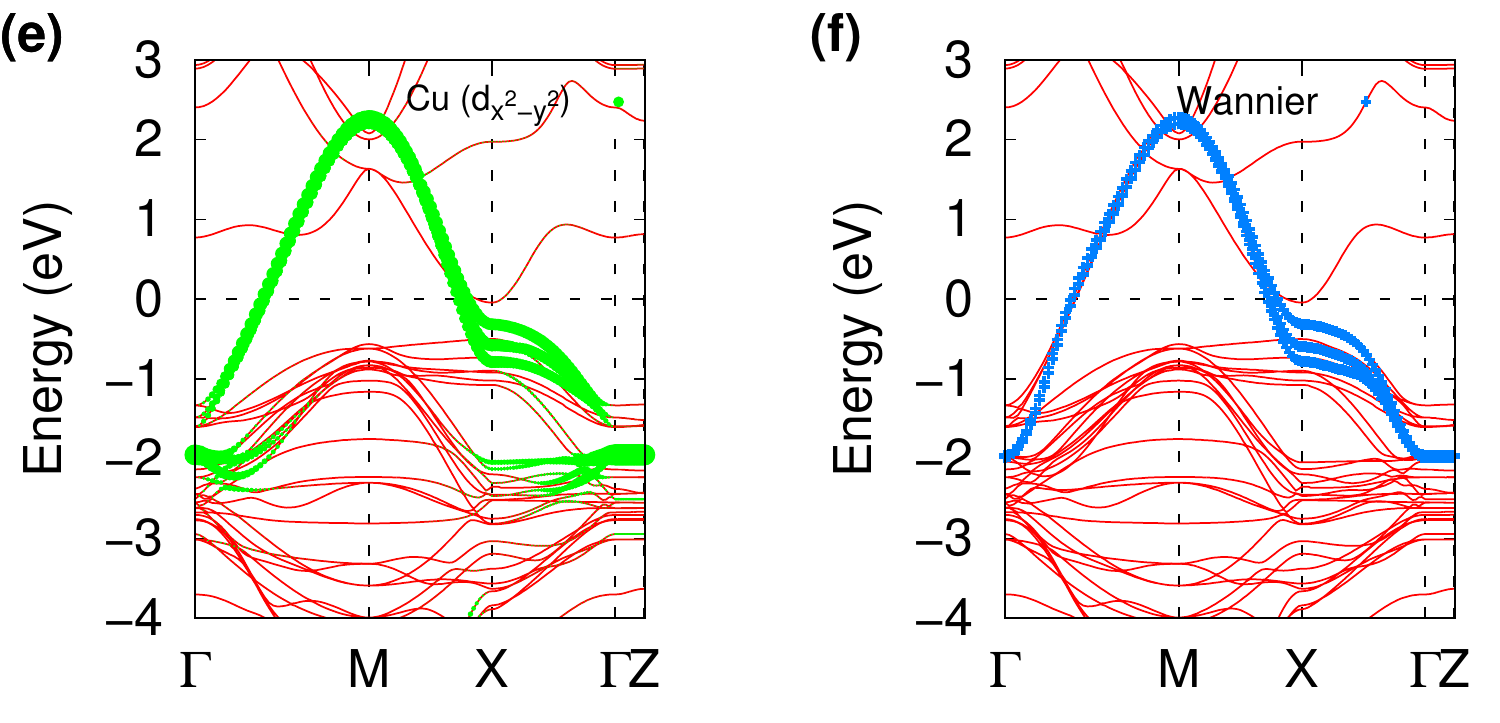}
\caption{ (Color online) The calculated LDA band structure (a, c, e) and the Wannier
  band (b, d, f) for Hg-compounds. The single-, double-, and
  triple-layer cases are presented in (a, b), (c, d), and (e, f),
  respectively. In (a, c, e), the Cu-$d_{x^2-y^2}$ character is
  represented by the green color. In (b, d, f), the $d_{x^2-y^2}$ Wannier band is
  depicted by the blue cross on top of the calculated band dispersion
  (red lines).}
\label{fig:Hg_Wannier}
\end{center}
\end{figure}

\begin{figure}[!t]
\begin{center}
\renewcommand{\thefigure}{S\arabic{figure}}
\includegraphics[width=12cm,angle=0]{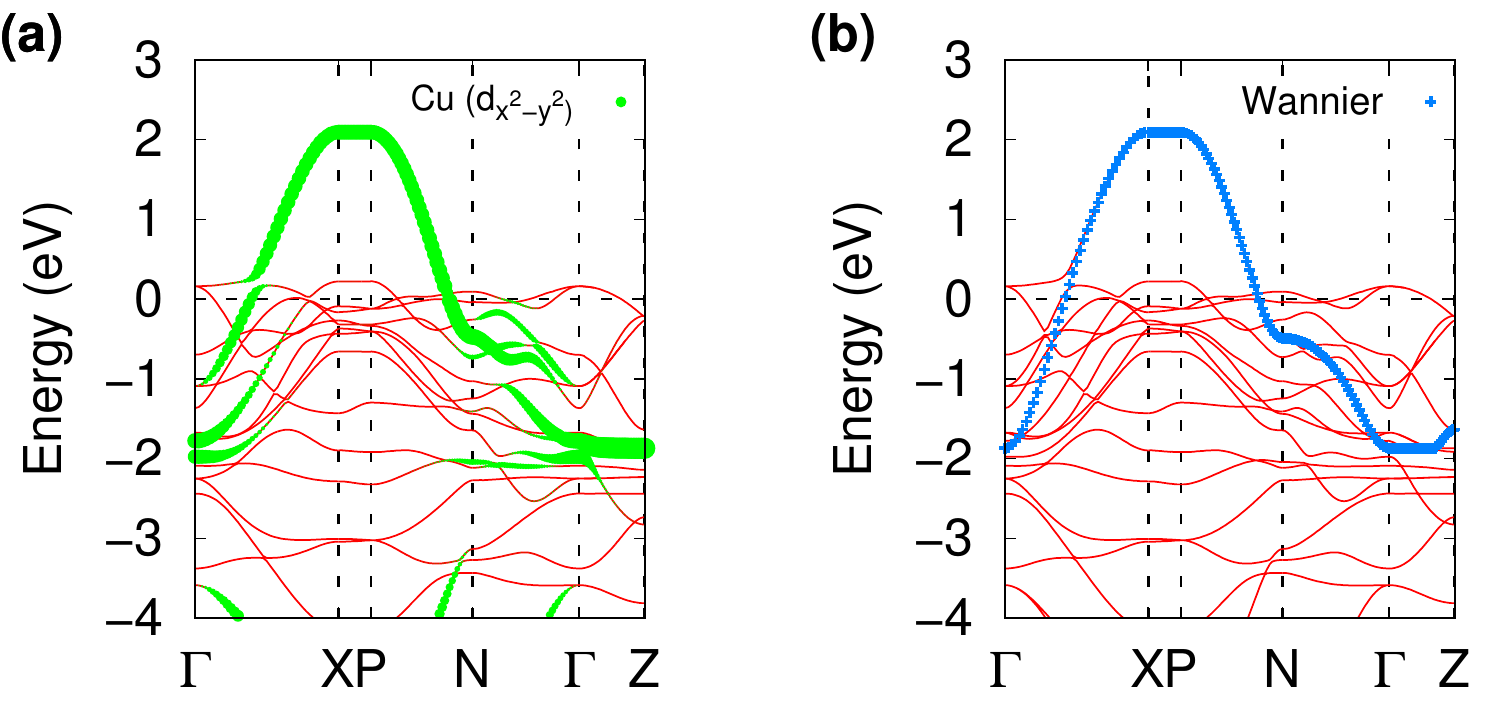}
\includegraphics[width=12cm,angle=0]{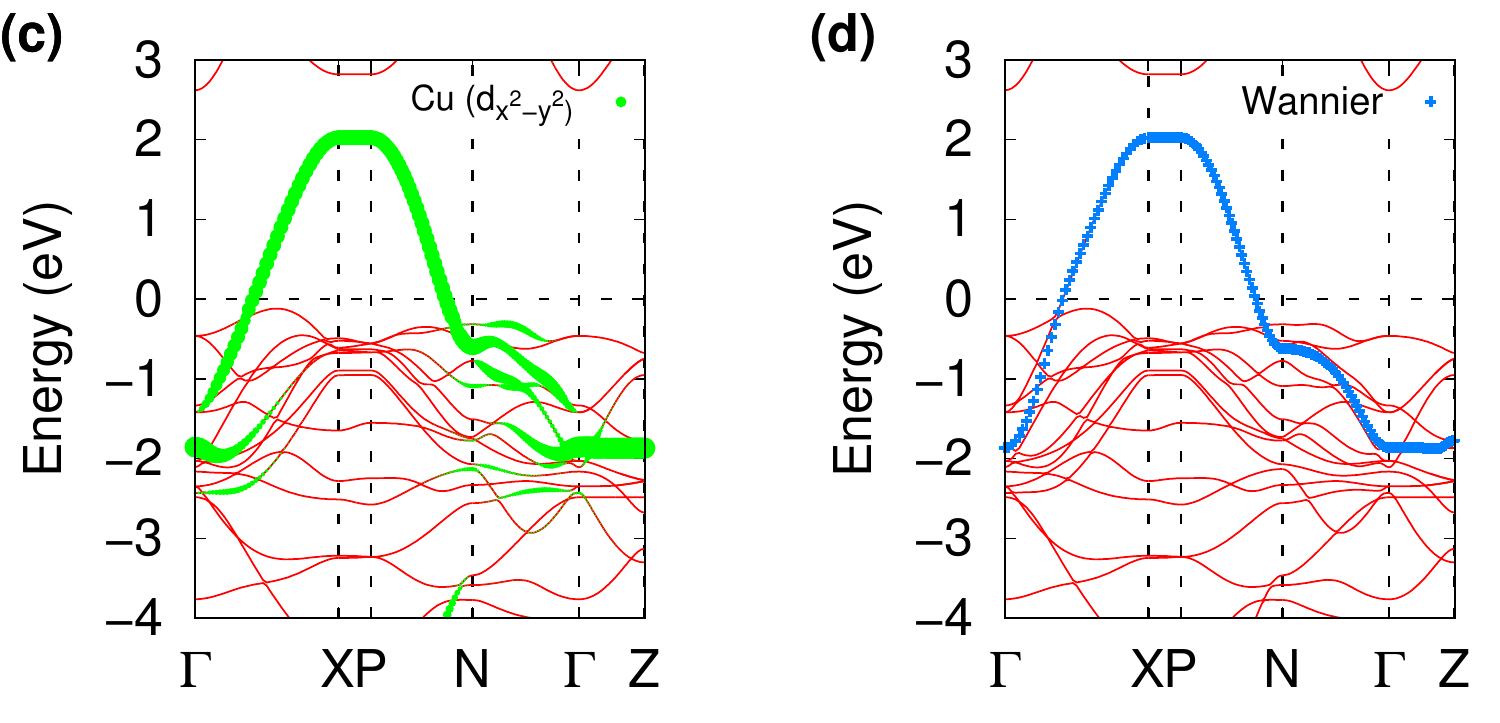}
\includegraphics[width=12cm,angle=0]{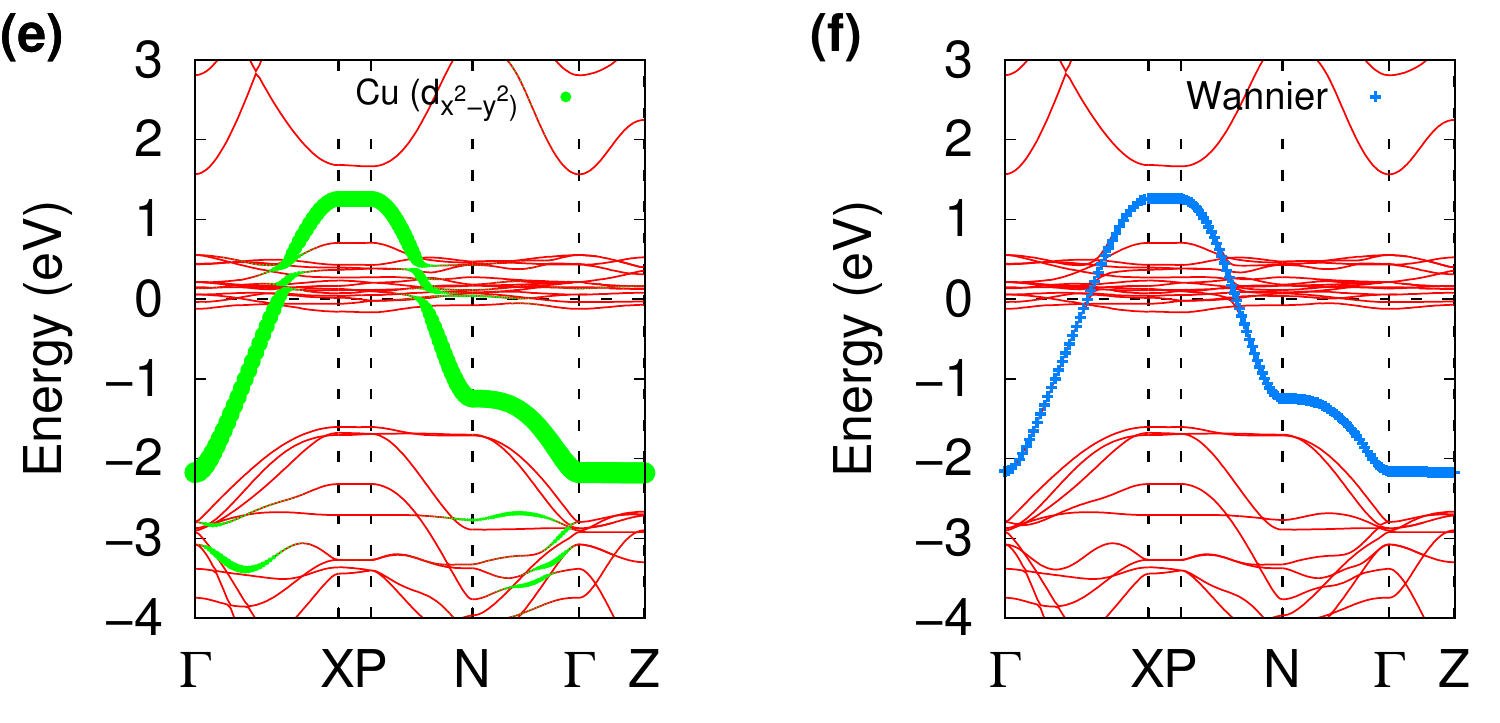}
\caption{ (Color online) The calculated LDA band structure (a, c, e) and the Wannier
  band (b, d, f) of Pr$_2$CuO$_4$. The results from Method 1, 2, and 3
  are presented in (a, b), (c, d), and (e, f), respectively. In (a, c,
  e), the Cu-$d_{x^2-y^2}$ character is represented by the green
  color. In (b, d, f), the $d_{x^2-y^2}$ Wannier band is depicted by the blue cross
  on top of the calculated band dispersion (red lines).}
\label{fig:Pr2CuO4_Wannier}
\end{center}
\end{figure}

We have performed the Wannier fit by considering both the $d_{x^2-y^2}$ and
$d_{z^2}$ orbitals explicitly as the internal space \cite{Sakakibara-PRL2010},
and here we present only the $d_{x^2-y^2}$ bands, for which U was estimated.
The Wannier band fitting works well for all cases including the
multilayer Hg-cuprates (having multiple Cu-$d_{x^2-y^2}$ bands) and
the RE$_2$CuO$_4$ (having RE-$4f$ states around Fermi energy; Method
3). The results are presented in Figure~\ref{fig:Hg_Wannier}
and~\ref{fig:Pr2CuO4_Wannier}.

\section{Model parameters}
The calculation results for $t$, $t'$, $U$, $U$/$t$, $\Delta_{dp}$, and $J$ are
summarized in Table~\ref{tab:Two-band model}. The
$\Delta_{dp}$ calculated by Weber {\it et al.} \cite{Weber-PRB2010} is
also presented for comparison.

\begin{table*}[!ht]
\renewcommand{\thetable}{S\arabic{table}}
\vspace{0.3cm}
\begin{ruledtabular}
\begin{tabular}{cccccccc}
				 	 & $t$ & $t'$ & $U$ & $U$/$t$ & $\Delta_{dp}$ & $\Delta_{dp}$ (Ref.\cite{Weber-PRB2010}) & $J$  \\[-0.25cm]
					Material 	 & (eV) & (eV) & (eV) & & (eV) & (eV) & (eV)	\\\hline					
	La$_2$CuO$_4$ 			  &  0.478 & 0.096 & 3.15 & 6.59 & 2.58 & 2.76 & 0.43 \\
	HgBa$_2$CuO$_4$ 			 & 0.476 & 0.095 & 2.15 &  4.52 & 1.84 & N/A & 0.58 \\
	HgBa$_2$CaCu$_2$O$_6$ 			  & 0.479 & 0.106 & 2.15 & 4.49 & 1.87 & N/A & 0.49 \\
	HgBa$_2$Ca$_2$Cu$_3$O$_8$	  & 0.482 & 0.106 & 1.48 & 3.07 & 1.95 & N/A & 0.25 \\ [-0.2cm]
	(Outer-layer) 	\\
		HgBa$_2$Ca$_2$Cu$_3$O$_8$			  & 0.486 & 0.107 & 1.17 & 2.41 & 1.88   & N/A & 0.33 \\ [-0.2cm]
	 (Inner-layer) 			 \\
	Pr$_2$CuO$_4$ (Method 1)			 &  0.483 & 0.088 & 0.87 & 1.80 & 1.37 & 1.65 & 0.39 \\
	Pr$_2$CuO$_4$ (Method 2) 			 &  0.462 & 0.090 & 1.72 & 3.72 & 1.66 & N/A & 0.52\\
	Pr$_2$CuO$_4$ (Method 3) 			 &  0.403 & 0.108 & 1.13 & 2.80 & 2.59 & N/A & 0.52\\
	Nd$_2$CuO$_4$ (Method 1)			 &  0.493 & 0.088 & 0.90 & 1.83 & 1.39 & 1.61 & 0.35\\
	Nd$_2$CuO$_4$ (Method 2) 			 &  0.470 & 0.093 & 1.82 & 3.87 & 1.72 & N/A & 0.52\\
	Nd$_2$CuO$_4$ (Method 3) 			 & 0.417 & 0.109 & 1.16 & 2.78 & 2.54 & N/A & 0.53 \\
	Sm$_2$CuO$_4$ (Method 1)			 & 0.491 & 0.091 & 0.81 & 1.65 & 1.42 & N/A & 0.31\\
	Sm$_2$CuO$_4$ (Method 2) 			  & 0.478 & 0.095 & 1.90 & 3.97 & 1.76 & N/A & 0.52 \\
	Sm$_2$CuO$_4$ (Method 3) 			  & 0.426 & 0.110 & 1.30 & 3.05 & 2.43 & N/A & 0.52\\
\end{tabular}
\end{ruledtabular}
\caption{The summary of the calculated model parameters.}
\label{tab:Two-band model}
\end{table*}